\documentclass[aps,showpacs,preprint,superscriptaddress]{revtex4}
\usepackage{graphicx}
\usepackage{subfigure}
\usepackage{float}
\usepackage{amsmath}
\usepackage{amsfonts}
\usepackage{bm}
\usepackage{bbm}
\usepackage{txfonts}
\usepackage{array}
\usepackage{amssymb}
\usepackage{color}
\begin{document}

\title{Asymmetric pulse effects on pair production in polarized electric fields}
\author{Obulkasim Olugh}
\affiliation{Key Laboratory of Beam Technology of the Ministry of Education, and College of Nuclear Science and Technology, Beijing Normal University, Beijing 100875, China}
\affiliation{Xinjiang Police College, Urumqi 830011, China}
\author{Zi-Liang Li}
\affiliation{School of Science, China University of Mining and Technology, Beijing 100083, China}
\author{Bai-Song Xie \footnote{Corresponding author. Email address: bsxie@bnu.edu.cn}}
\affiliation{Key Laboratory of Beam Technology of the Ministry of Education, and College of Nuclear Science and Technology, Beijing Normal University, Beijing 100875, China}
\affiliation{Beijing Radiation Center, Beijing 100875, China}
\date{\today}
\begin{abstract}
Using the Dirac-Heisenberg-Wigner (DHW) formalism, effects of asymmetric pulse shape on the generation of electron-positron pairs in three typical polarized fields, i.e., the linear, middle elliptical and circular ones, are investigated. Two kinds of asymmetries for the falling pulse length, one is compressed and the other is elongated, are studied. It is found that the interference effect disappears with the compression of the pulse length, and finally the peak value of the momentum spectrum is concentrated in the center of the momentum space. For the opposite situation by extending the falling pulse length, a multi-ring structure without interference appears in the momentum spectrum. Research results exhibit that the momentum spectrum is very sensitive to the asymmetry of the pulse as well as to the polarization of the fields. It is also found that the number density of electron-positron pairs under different polarizations is sensitive to the asymmetry of electric field. For the compressed falling pulse, the number density can be enhanced significantly over $2$ orders of magnitude. These results could be useful in planning high power or/and high-intensity laser experiments.
\end{abstract}
\maketitle

\section{Introduction}

In intense electromagnetic fields the vacuum state is unstable and spontaneously
to generate electron-positron pairs. This is known as the Schwinger effect, which is one of the highly nontrivial predictions in quantum electrodynamics (QED) \cite{Sauter:1931zz,Heisenberg:1935qt,Schwinger:1951nm}.
Due to the tunneling nature of the Schwinger effect, this interesting phenomenon is exponentially suppressed and the pair production rate is proportional to $\exp (-\pi E_{cr} /E )$, where the corresponding Schwinger critical field strength $E_{cr} =  {m_e^2c^3} / {e\hbar} = 1.3 \times 10^{18}\rm{V/m}$. The associated laser intensity, {\it e.g.},  $I=4.3 \times10^{29}$W/cm$^{2}$, is too high and beyond current technological possibilities. Therefore, its detection has remained a challenge for many decades\cite{Gelis:2015kya}. However, current advances in high-power laser technology \cite{Heinzl:2008an,Marklund:2008gj,Pike:2014wha} and the forthcoming available experiments (for example, in view of planned facilities as the Extreme Light Infrastructure (ELI), the Exawatt Center
for Extreme Light Studies (XCELS), or the Station of Extreme Light at the Shanghai
Coherent Light Source) have brought the hope the QED predictions enter the realm of observation. On the other hand, by using x-ray free electron laser (XFEL) facilities can in principle get a strong field at about~$E=0.1E_{cr}= 1.3 \times 10^{17}\rm{V/m}$ \cite{Ringwald:2001ib} and drive interest in studying pair production under superstrong fields.

Schwinger effect is one of the nonperturbative phenomena in QED, therefore studying the pair production in the nonperturbative regime would deepen our knowledge about the relatively less tested branch of QED. Motivated by this, many exploratory studies of the Schwinger effect based on a number of different theoretical techniques have been undertaken, for example, within the quantum kinetic approach \cite{Alkofer:2001ik,Roberts:2002py} and the real time Dirac-Heisenberg-Wigner (DHW) formalism \cite{Vasak:1987um,Hebenstreit:2011pm,Kohlfurst:2015zxi}, WKB approximation \cite{Akkermans:2011yn,Dumlu:2010ua} as well as worldline instanton technique \cite{Schutzhold:2008pz}. In \cite{Hebenstreit:2009km} by using the quantum kinetic approach the momentum spectrum of the produced particles has been computed, and spectrum was found to be extremely sensitive to these
physical pulse parameters. The concrete description for various approaches in detail and some latest publications can be found in our recent review for pair production \cite{MRE2017}.

In this paper, we shall further investigated the Schwinger effect by considering asymmetric pulse shape with Gaussian envelope and different polarizations.
We mainly consider asymmetric pulse shape effects on pair production in different polarization, {\it e.g.}, linear polarization, elliptic polarization and circular polarization. We will reveal some novel features of the momentum spectra of created pairs for differently polarized electric fields. In this study the
real-time Dirac-Heisenberg-Wigner~(DHW)~formalism is employed. Because the DHW formalism is very efficient for the calculation of involving circularly \cite{Blinne:2013via,Blinne:2016yzv} or elliptically polarized electric fields \cite{Li:2015cea,olugh1}. 

This manuscript is organized as follows. In Sec.\ref{result2}, we introduce the model of a background field. In Sec.\ref{method}, we introduce briefly the DHW formalism which is used in our calculation for completeness. In Sec.\ref{result3}, we show the numerical results for momentum spectra and analyze the underlying physics. In Sec.\ref{number}, we give the numerical results for the pair number density. We end up the paper with a brief summary and discussion in the last section.

\section{external electric field model}\label{result2}

We focus on the study of $e^{-}e^{+}$ pair production in differently polarized and time dependent asymmetric electric fields.
So the explicit form of the external field is given as
\begin{equation}\label{eq1}
\mathbf{E}(t) \,\, =\, \, \frac{E_{0}}{\sqrt{1+\delta^{2}}}\,
\left(e^{-\frac{1}{2}(t/\tau_{1})^{2}}\theta(-t)+e^{-\frac{1}{2}(t/\tau_{2})^{2}}\theta(t)\right) \,
\left(\begin{array}{c}
          \cos(\omega t+\phi) \\
           \delta\sin(\omega t+\phi) \\
              0 \\
        \end{array}\right),
\end{equation}
where $\frac{E_{0}}{\sqrt{1+\delta^{2}}}$ for the field amplitudes, $\tau_{1}$ and $\tau_{2}$ are the rising and falling pulse durations, respectively, $\theta(t)$ is the Heaviside step function, and $\omega$ the
oscillation frequency, $\phi$ is the carrier phase, and $\mid\delta\mid\leq1$ represents the field polarization (or the ellipticity).
The field parameters chosen as: $E_{0}=0.1\sqrt{2}E_{cr}$, $\omega=0.6m$, and $\tau_{1}=10/m$, $\phi=0$, where $m$ is the electron mass.
For the falling pulse length we set the parameter as $\tau_{2}=k\tau_{1}$, where $k$ is the ratio of the falling to rising pulse length. Throughout this paper we use natural units $\hbar=c=1$.

The main interest in this study is asymmetric pulse durations effects on pair production in differently polarized and time dependent asymmetric electric fields. We mainly consider two different situations. One is that the rising pulse length $\tau_{1}$ is fixed
and the falling pulse length $\tau_{2}=k\tau_{1}$ becomes shorter with $0<k\leq1$. The other is that when the rising pulse length $\tau_{1}$ is fixed and the falling pulse length $\tau_{2}=k\tau_{1}$ becomes longer with $k\geq 1$.

\section{A brief outline on DHW formalism}\label{method}

The DHW formalism is an approach to describe the quantum phenomena of a system by a Wigner function as the relativistic phase space distribution that has many advantages and practical usages. It has been also further adopted in the studies of Sauter-Schwinger QED vacuum pair production \cite{Vasak:1987um,Hebenstreit:2011pm}.
In the following, we present a brief outline to DHW formalism for a completeness of paper self-containing.

A convenient starting point is the gauge-invariant density operator of two Dirac field operators in the Heisenberg picture
\begin{equation}\label{density}
 \hat {\mathcal C}_{\alpha \beta} \left( r , s \right) = \mathcal U \left(A,r,s
\right) \ \left[ \bar \psi_\beta \left( r - s/2 \right), \psi_\alpha \left( r +
s/2 \right) \right],
\end{equation}
in terms of the electron's spinor-valued Dirac field $\psi_\alpha (x)$, where
$r$ denotes the center-of-mass and $s$ the relative coordinates, respectively. The Wilson-line factor before the commutators
\begin{equation}
 \mathcal U \left(A,r,s \right) = \exp \left( \mathrm{i} e s \int_{-1/2}^{1/2} d
\xi \ A \left(r+ \xi s \right)  \right)
\end{equation}
is used to keep the density operator gauge-invariant, and this factor depends on the
elementary charge $e$ and the background gauge field $A$, respectively.  In addition, we use a meanfield (Hartree) approximation via replacing the gauge field operator with the background field.

The important quantity of the DHW  method is the covariant Wigner operator given as the Fourier transform of the density operator \eqref{density}
\begin{equation}
 \hat{\mathcal W}_{\alpha \beta} \left( r , p \right) = \frac{1}{2} \int d^4 s \
\mathrm{e}^{\mathrm{i} ps} \  \hat{\mathcal C}_{\alpha \beta} \left( r , s
\right).
\end{equation}
By taking the vacuum expectation value of the Wigner operator, it gives the Wigner function as
\begin{equation}
 \mathbbm{W} \left( r,p \right) = \langle \Phi \vert \hat{\mathcal W} \left( r,p
\right) \vert \Phi \rangle.
\end{equation}
By decomposing the Wigner function in terms of a complete basis set 
of Dirac matrices, we can get 16 covariant real Wigner components
\begin{equation}
\mathbbm{W} = \frac{1}{4} \left( \mathbbm{1} \mathbbm{S} + \textrm{i} \gamma_5
\mathbbm{P} + \gamma^{\mu} \mathbbm{V}_{\mu} + \gamma^{\mu} \gamma_5
\mathbbm{A}_{\mu} + \sigma^{\mu \nu} \mathbbm{T}_{\mu \nu} \right) \, .
\label{decomp}
\end{equation}
According to the  Ref. \cite{Vasak:1987um,Hebenstreit:2011pm} the equations of motion for the Wigner function are
\begin{equation}
D_{t}\mathbbm{W} = -\frac{1}{2}\mathbf{D}_{\mathbf{x}}[\gamma^{0}\bm{\gamma},\mathbbm{W}]
+im[\gamma^{0},\mathbbm{W}]-i\mathbf{P}\{\gamma^{0}\bm{\gamma},\mathbbm{W}\},
\label{motion}
\end{equation}
where $D_{t}$, $\mathbf{D}_{\mathbf{x}}$ and $\mathbf{P}$ denote the pseudodifferential operators
\begin{equation}
\begin{array}{l}
D_{t}=\partial_{t}+e\int^{1/2}_{-1/2}d\lambda\, \mathbf{E}(\mathbf{x}+i\lambda\mathbf{\nabla}_{\mathbf{p}},t)\cdot\mathbf{\nabla}_{\mathbf{p}},\\
\mathbf{D}_{\mathbf{x}}=\mathbf{\nabla}_{\mathbf{x}}+e\int^{1/2}_{-1/2}d\lambda\, \mathbf{B}(\mathbf{x}+i\lambda\mathbf{\nabla}_{\mathbf{p}},t)\times\mathbf{\nabla}_{\mathbf{p}},\\
\mathbf{P}=\mathbf{p}-ie\int^{1/2}_{-1/2}d\lambda\, \lambda\,\mathbf{B}(\mathbf{x}+i\lambda\mathbf{\nabla}_{\mathbf{p}},t)\times\mathbf{\nabla}_{\mathbf{p}}.
\end{array}\label{eq2}
\end{equation}

Inserting the decomposition Eq. \eqref{decomp} into the equation of motion Eq. \eqref{motion} for the Wigner function, one can obtain a set of partial differential equations (PDEs) for the 16 Wigner components.
Furthermore, for the spatially homogeneous electric fields like Eq. \eqref{eq1}, by using the characteristic method \cite{Blinne:2013via}, replacing the kinetic momentum ${\mathbf p}$ with the canonical momentum ${\mathbf q}$ via $ {\mathbf q} - e {\mathbf A} (t)$, and the PDEs for the 16 Wigner components can be reduced to 10 ordinary differential equations(ODEs) of the nonvanishing Wigner coefficients
\begin{equation}
{\mathbbm w} = ( {\mathbbm s},{\mathbbm v}_i,{\mathbbm a}_i,{\mathbbm t}_i)
\, , \quad  {\mathbbm t}_i := {\mathbbm t}_{0i} -   {\mathbbm t}_{i0}  \, .
\end{equation}
For the detailed derivations and explicit form of the 10 equations, one can refer to the
Refs. \cite{Hebenstreit:2011pm,Kohlfurst:2015zxi,olugh}. By the way the corresponding vacuum nonvanishing initial values are
\begin{equation}
{\mathbbm s}_{vac} = \frac{-2m}{\sqrt{{\mathbf p}^2+m^2}} \, ,
\quad  {\mathbbm v}_{i,vac} = \frac{-2{ p_i} }{\sqrt{{\mathbf p}^2+m^2}} \, .
\end{equation}

In the following, one can expresses the scalar Wigner coefficient
by the one-particle momentum distribution function
\begin{equation}
f({\mathbf q},t) = \frac 1 {2 \Omega(\mathbf{q},t)} (\varepsilon - \varepsilon_{vac} ),
\end{equation}
where $\Omega(\mathbf{q},t)= \sqrt{{\mathbf p}^2(t)+m^2}=
\sqrt{m^{2}+(\mathbf{q}-e\mathbf{A}(t))^{2}}$ is the total energy of the electron's (positron's) and
$\varepsilon = m {\mathbbm s} + p_i {\mathbbm v}_i$
is the phase-space energy density.
To obtain one-particle momentum distribution function $f({\mathbf q},t)$, referring to \cite{Blinne:2013via}, it is helpful to introduce an auxiliary three-dimensional vector $\mathbf{v}(\mathbf{q},t)$
\begin{equation}
v_i (\mathbf{q},t) : = {\mathbbm v}_i (\mathbf{p}(t),t) -
(1-f({\mathbf q},t))  {\mathbbm v}_{i,vac} (\mathbf{p}(t),t).
\end{equation}
So the one-particle momentum distribution function $f({\mathbf q},t)$ can be obtained by solving the following ordinary differential equations including it as well as the other nine auxiliary quantities,
\begin{equation}
\begin{array}{l}
\dot{f}=\frac{e\mathbf{E}\cdot \mathbf{v}}{2\Omega},\\
\dot{\mathbf{v}}=\frac{2}{\Omega^{3}}[(e\mathbf{E}\cdot \mathbf{p})\mathbf{p}-e\mathbf{E}\Omega^{2}](f-1)-\frac{(e\mathbf{E}\cdot \mathbf{v})\mathbf{p}}{\Omega^{2}}-2\mathbf{p}\times \mathbf{a}-2m\mathbf{t},\\
\dot{\mathbf{a}}=-2\mathbf{p}\times \mathbf{v},\\
\dot{\mathbf{t}}=\frac{2}{m}[m^{2}\mathbf{v}-(\mathbf{p}\cdot \mathbf{v})\mathbf{p}],
\end{array}\label{eq3}
\end{equation}
with the initial conditions $f(\mathbf{q},-\infty)=0$, $\mathbf{v}(\mathbf{q},-\infty)=\mathbf{a}(\mathbf{q},-\infty)=\mathbf{t}(\mathbf{q},-\infty)=0$, where the time derivative is indicated by a dot, $\mathbf{a}(\mathbf{q},t)$ and $\mathbf{t}(\mathbf{q},t)$ are the three-dimensional vectors corresponding to Wigner components, and $\mathbf{A}(t)$ denotes the vector potential of the external field.

Finally, by integrating the distribution function $f(\mathbf{q},t)$ over full momentum space, we obtain the number density of created pairs defined at asymptotic times $t\rightarrow+\infty$:
\begin{equation}\label{14}
  n = \lim_{t\to +\infty}\int\frac{d^{3}q}{(2\pi)^ 3}f(\mathbf{q},t) \, .
\end{equation}

\section{Momenta spectra of the produced particles}\label{result3}

In this section, we will report some interesting results for the momenta spectra of the
produced particles with several pulse parameters under typical cases of polarization field such as
linear ($\delta=0$), elliptical ($\delta=0.5$) and circular ($\delta=1$) ones.

\subsection{Linear polarization $\delta=0$}\label{result3.1}

\begin{figure}[htbp]
\suppressfloats
\begin{center}
\includegraphics[width=\textwidth]{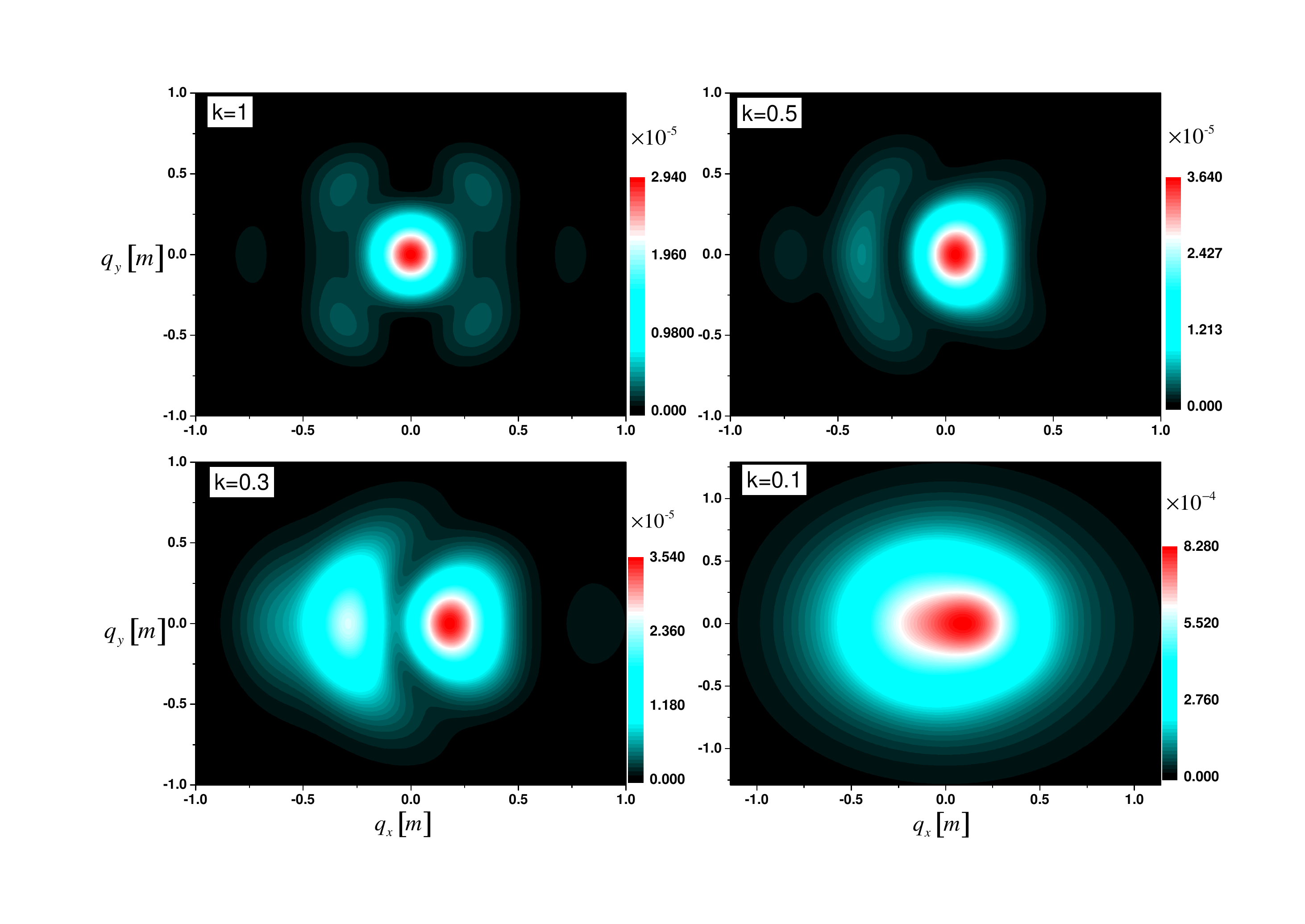}
\end{center}
\caption{Momentum spectra of produced $e^{+}e^{-}$ pairs for linear polarization ($\delta=0$)
at $q_z=0$ in the $( q _x,q_y)$-plane when the rising pulse length $\tau_{1}$ is fixed but
the falling pulse length $\tau_{2}=k\tau_{1}$ becomes shorter with $0<k\leq1$. The chosen parameters are $E_{0}=0.1\sqrt{2}E_{cr}$, $\omega=0.6m$, and $\tau_{1}=10/m$, where $m$ is the electron mass.}
\label{fig:1}
\end{figure}

Firstly, when one keeps the rising pulse length $\tau_{1}$ fixed but
changes the falling pulse length $\tau_{2}=k\tau_{1}$ to be shorter with $0<k\leq1$, the momentum spectra are shown in Fig. \ref{fig:1} for different $k$.

\begin{figure}[htbp]
\begin{center}
\includegraphics[width=\textwidth]{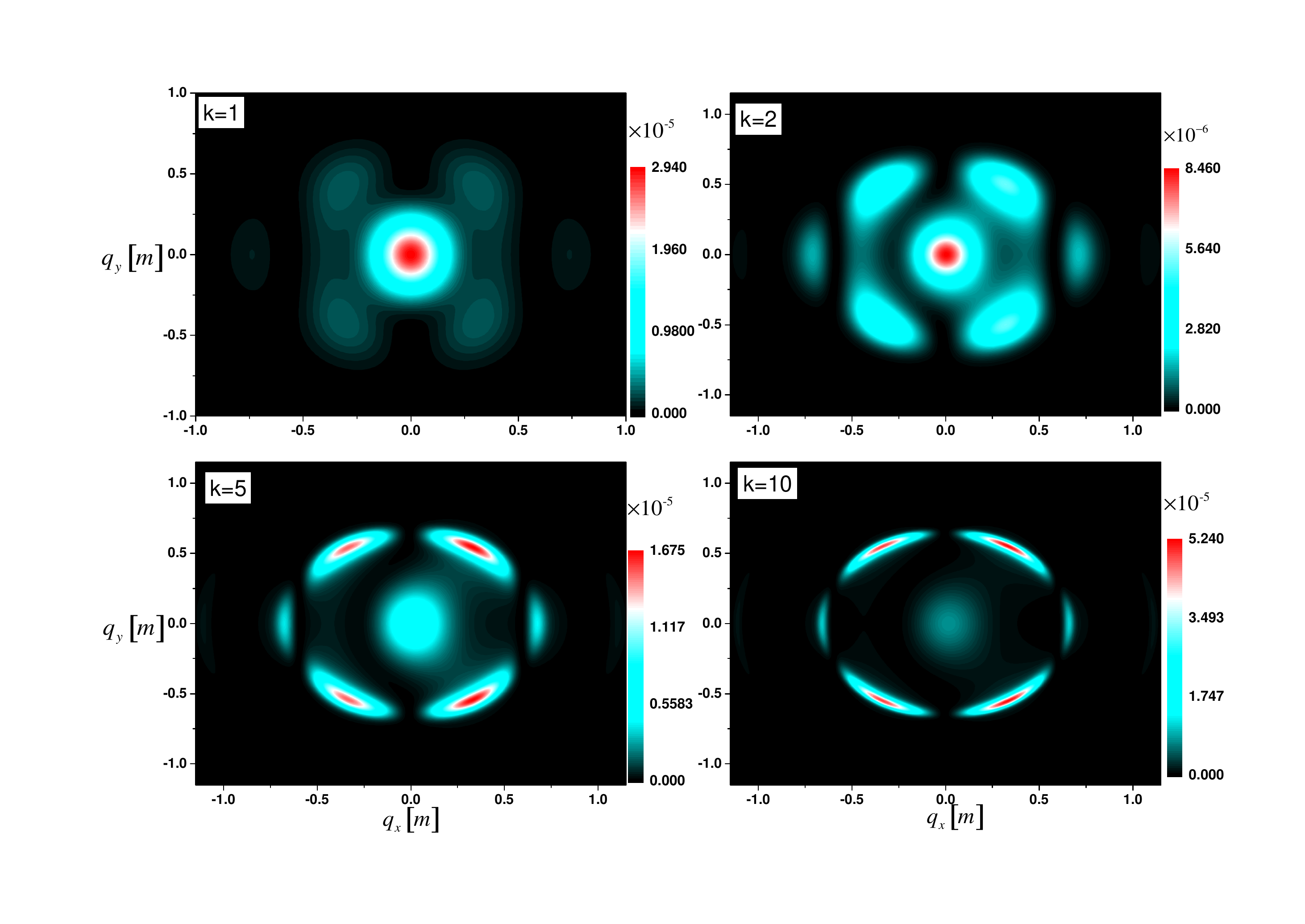}
\end{center}
\caption{Same as in Fig.\ref{fig:1} except the falling pulse length $\tau_{2}=k\tau_{1}$ becomes longer with $k\geq1$.}
\label{fig:2}
\end{figure}

For $k=1$ the momentum spectrum is centered at the origin, and weak oscillation is observed, as shown in the upper-left in Fig. \ref{fig:1}. The physical origin of the oscillation is explained in \cite{Dumlu:2010ua} in terms of the interference between separate complex conjugate pairs of turning points.

Interesting findings can be seen that the momentum spectrum of the created pair is very sensitive to the asymmetry of the electric field. When the ratio parameter $k$ is changed to $k=0.5$, the main peak of the momentum spectrum is shifted to the positive $q_{x}$ and the symmetry distribution of the momentum spectrum is destroyed. This effect is similar to the effect of carrier phase studied in \cite{Hebenstreit:2009km}. Furthermore when $k=0.3$, the main parts of momentum spectrum appear also to the negative $q_{x}$ beside the positive $q_{x}$ peak, which means the split of momentum spectrum. Therefore, two peaks are observed. This result is similar to the effect introduced by the frequencies chirp in \cite{olugh}. For the very asymmetric case of $k=0.1$, the momentum spectrum of the particle is concentrated again in the center but the oscillation of the momentum spectrum disappears. Finally, it is noted that the peak value of the momentum spectrum of the pairs is increased from $2.94\times10^{-5}$ ($k=0$) to $8.28\times10^{-4}$ ($k=0.1$).

Secondly, when the rising pulse length $\tau_{1}$ is fixed but the falling pulse length $\tau_{2}$ becomes longer with $k\geq 1$, the result of momentum spectrum are shown in Fig. \ref{fig:2}.
From these figures, we can see that as field asymmetry increases, the main center peak of momentum spectrum decreases while some disconnected ring-like structures with peaks appear and gradually become main ones. And this tendency is more striking with larger $k$.

In detail we find that the center maximum value of momentum spectrum decreases until $k\leq5$. For the pulse length increases to $k=10$, the maximum value at ring is larger slightly than that of symmetrical pulse when $k=1$. Note that the ring structure in the momentum spectrum is the typical features of the multiphoton pair production mechanism. For example, the inner ring is formed by absorbing four photons, and the outermost obscured structure corresponds to the absorption of five photons.

\subsection{Elliptic polarization $\delta=0.5$}\label{result3.2}

\begin{figure}[htbp]
\suppressfloats
\begin{center}
\includegraphics[width=\textwidth]{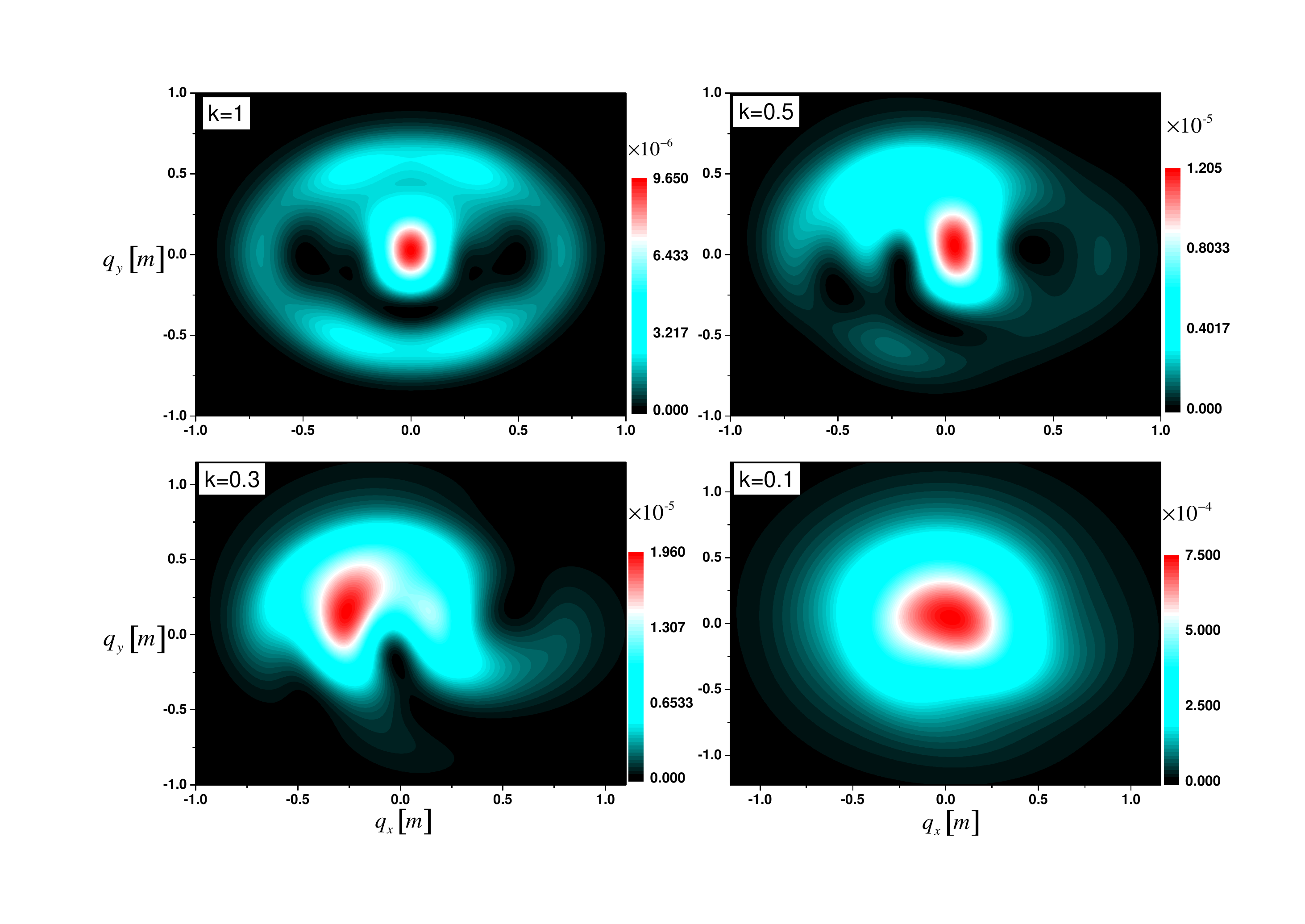}
\end{center}
\caption{Same as in Fig.\ref{fig:1} except for elliptic polarization $\delta=0.5$.}
\label{fig:3}
\end{figure}

For middle-elliptical polarization case $\delta=0.5$, the result of momentum spectrum for compressed pulse cases are exhibited in Fig. \ref{fig:3}. From the top left of Fig.\ref{fig:3}, where $k=1$, one can see that the momentum spectrum is symmetrically distributed for $q_{x}$ axis, and spectrum peak is located at $\mathbf{q}=0$. With $k$ decreasing, we can observe that the distortion of the momentum spectrum occurs, equivalently, the mirror symmetry about $q_{x}$ is lost. As the peak position is shifted, the maximum value of peak is increased. For example, when $k=0.5$, the main peak shifts along the positive $q_{y}$ direction while when $k=0.3$, the main peak shifts along the negative $q_{x}$ direction with a little larger peak value. For very asymmetric
case of $k=0.1$, the momentum spectrum is concentrated in the surrounding of the center and the main peak almost locates at the center again.

\begin{figure}[htbp]
\begin{center}
\includegraphics[width=\textwidth]{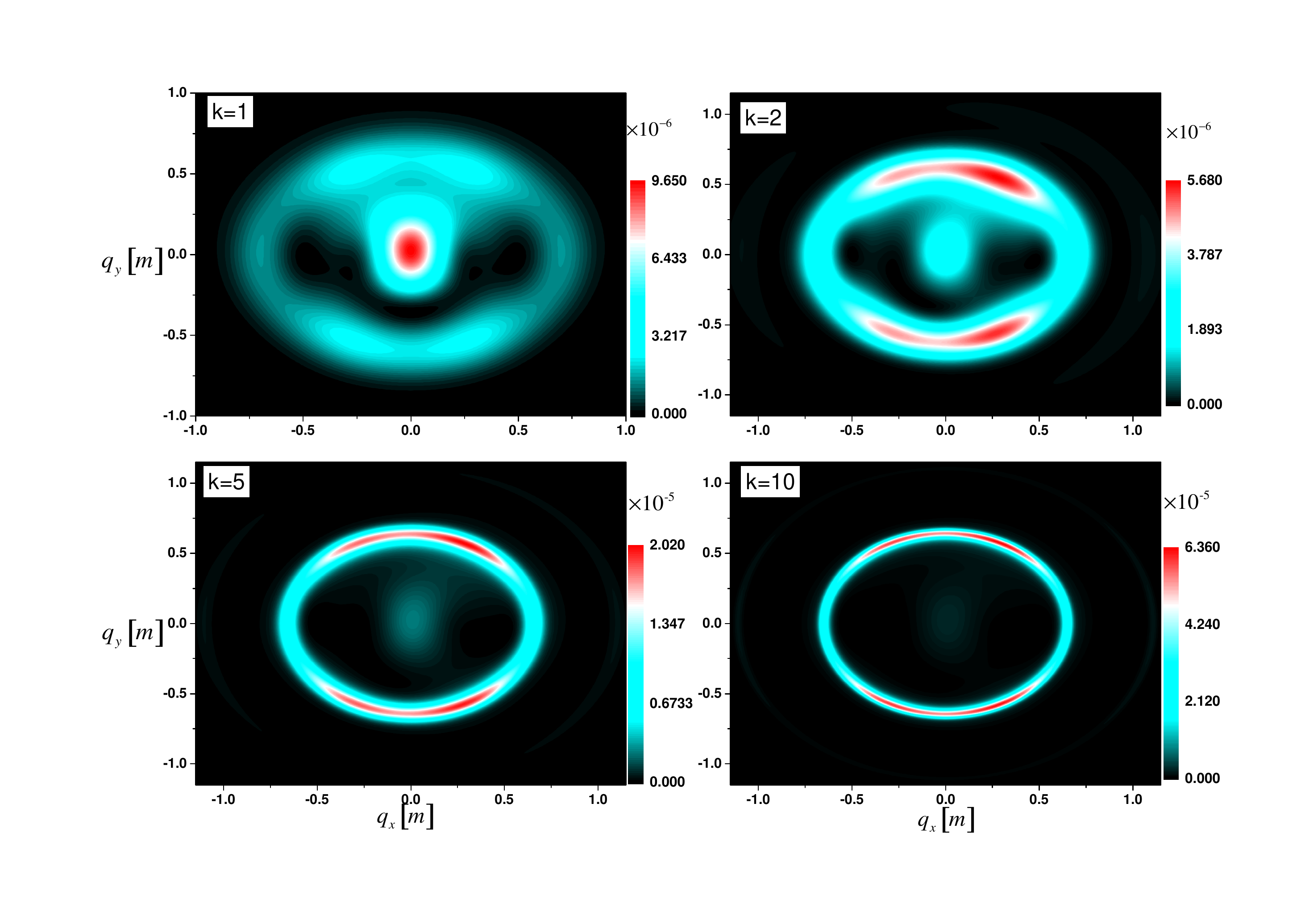}
\end{center}
\caption{Same as in Fig.\ref{fig:2} except for elliptic polarization $\delta=0.5$.}
\label{fig:4}
\end{figure}

Now let us consider the elongated falling pulse cases with $k\geq1$ for middle-elliptical polarization case $\delta=0.5$, the results of momentum spectrum are displayed in Fig. \ref{fig:4}. For $k=2$, the symmetry of momentum spectrum about the $q_{x}$ is destroyed. The peak position shifts to the positive and negative $q_{y}$ direction, while the peak value decreases compared to the symmetric case $k=1$. For the larger $k$, the spectrum at the center vanishes gradually with $k$ increasing, and a complete ring-like shape appears. The peaks positions are very interesting which form two elongated strips by locating at the relative narrower regime of positive and negative $q_{y}$ but relative broader regime of positive and negative $q_{x}$. Finally the additional outer ring structure appears again which is a clear signal for multiphoton pair production processes.
This can be understood from the fact that with the increases of pulse length $k\tau_1$, the electric field has enough long duration and changes its direction during the pair creation process. Thus, the created particles may be accelerated into different directions depending on the field direction at the time of production. This results in ring structure of the spectrum. On the other hand, as the pulse duration increases with $k$, the number of oscillation cycles within the Gaussian envelope also increases, and there will be more photons contributing to pair production by multiphoton absorption mechanism, so the signal of multiphoton pair creation becomes pronounced. However, for the compressed pulse cases, the number of oscillation in the envelope is very small, which does not show the standard multiphoton pair production clearly (although for the small pulse length $\tau$, the Keldysh parameter will be $\gamma=m/eE_{0}\tau>1$, but it is not strictly a multiphoton process). This explanation is also appropriate for linear and circular polarization.

\subsection{Circular polarization $\delta=1$}\label{result3.3}

\begin{figure}[htbp]
\begin{center}
\includegraphics[width=\textwidth]{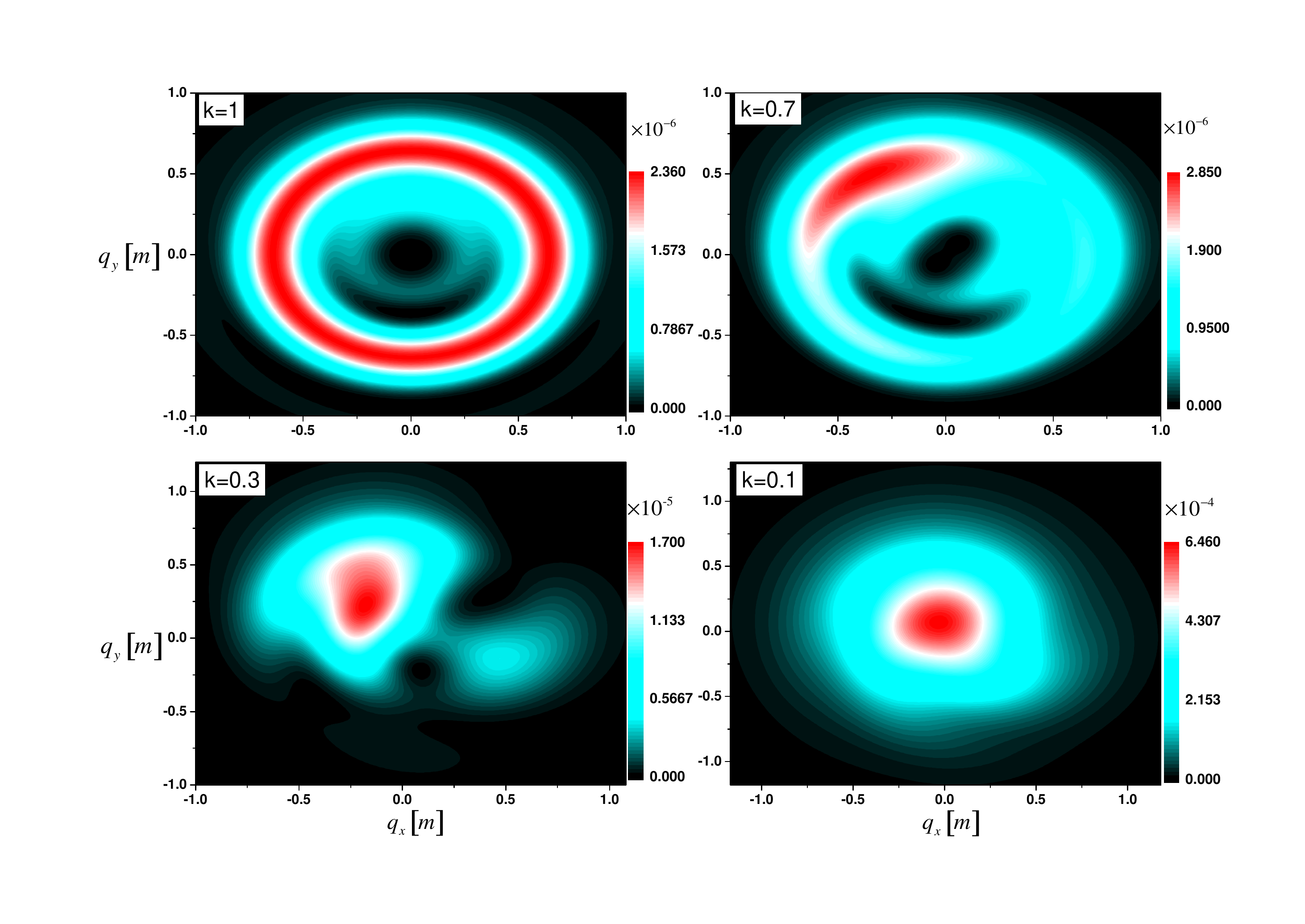}
\end{center}
\caption{Same as in Fig.\ref{fig:1} except for circular polarization $\delta=1$.}
\label{fig:5}
\end{figure}

For the circular polarization $\delta=1$, when the pulse length is compressed with $0<k\leq1$, the results of momentum spectrum are shown in Fig. \ref{fig:5}. From it one can see that in the symmetric case of $k=1$, the momentum spectrum has an obvious ring structure centered around the origin, meanwhile, a weak interference effect is also observed. The ring shape comes from absorbing four photons in the multiphoton pair production. We know that the ring radius can be calculated by the energy conservation by including the effective mass consideration, as $\mid \mathbf{q}\mid=1/2\sqrt{(n\omega)^{2}+(2m^{\ast})^{2}}$, where $n$ is the number of photons participating in the pair creation and $m^{\ast}$ is effective mass \cite{Kohlfurst:2013ura}. The weak interference effect can be explained by analysing the distribution of turning
points in semiclassical picture \cite{olugh}. The complex-valued turning points are those $t_{p}$ that are obtained by $\Omega(q,t_{p})=0$, which is responsible for the interference effects of the spectrum.

\begin{figure}[htbp]
\begin{center}
\includegraphics[width=\textwidth]{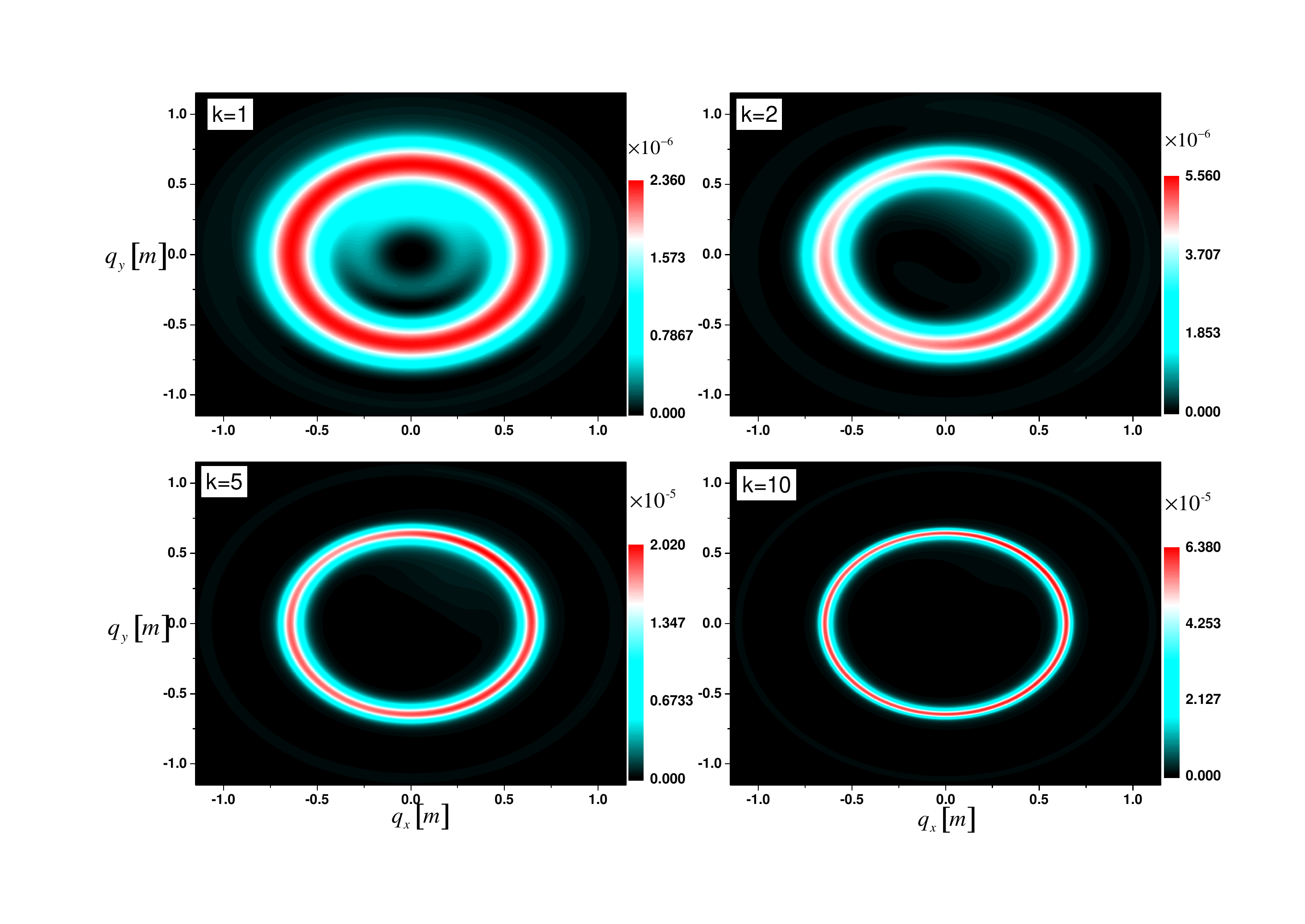}
\end{center}
\caption{Same as in Fig.\ref{fig:2} except for circular polarization $\delta=1$.}
\label{fig:6}
\end{figure}

With $k$ decreasing, the peaks of the momentum spectra display a quite rich structure and the interference effects vanish gradually. When $k=0.7$, the peak appears in the upper-left side of the momentum spectrum space. When $k=0.3$, the partial ring structure vanishes and the momentum spectrum becomes distorted. For the very asymmetric case of $k=0.1$, the peak position located at the near central region. Note that, for the circular polarization, the peak value of the momentum spectrum is enhanced remarkably by $2$ orders of magnitude compared to that in the symmetric case $k=1$.

We consider again the opposite situations for the falling pulse change, i.e., the falling pulse length $\tau_{2}$ becomes longer with $k \geq 1$. The result of momentum spectrum are shown in Fig. \ref{fig:6}.
It is obvious that, in this case, momentum distribution at the inner part of the ring vanishes gradually with $k$ increasing, and the red ring distribution becomes thin with a lacking of interference effect. Finally the additional outer ring shape appears again although it is a little obscure. The red inner ring in the momentum spectrum corresponds to $4$ photons absorbing, however, the outer ring is for absorbing $5$ photons.

\begin{table}[htbp]
\caption{The peak values of the particle distribution function, $f(\mathbf{q},\infty)$, for the typical polarization $\delta$ when the rising pulse length $\tau_{1}=10/m$ is fixed and the falling pulse length $\tau_{2}=k\tau_{1}$ is compressed or/and elongated. Note that these peaks occur at different values of the momentum $\mathbf{q}$.}
\centering
\begin{ruledtabular}
\begin{tabular}{cccccc}
$f_{max}(\mathbf{q},\infty)$ at peak & $(k= 1)$& $(k=0.1)$ & $(k=10)$ \\
\hline
$\delta=0$    &$29.40\times10^{-6}$ & $8.28\times10^{-4}$ & $5.24\times10^{-5}$\\
\hline
$\delta=0.5$ & $9.65\times10^{-6}$& $7.50\times10^{-4}$ & $6.36\times10^{-5}$\\
\hline
$\delta=1$    &$2.36\times10^{-6}$ & $6.46\times10^{-4}$ & $6.38\times10^{-5}$\\
\end{tabular}
\end{ruledtabular}
\vskip12pt
\end{table}

In Table \uppercase\expandafter{\romannumeral1} we list some corresponding peak values of momentum distribution for different polarization. It is found that, in the compressed cases of the falling pulse, the peak value of momentum spectrums is enhanced but this enhancement decreases as field polarization increases. In the vice versa case, i.e., the falling pulse is elongated, the peak value is enhanced also. However, on one hand, this enhancement increases as field polarization increases, on the other hand, the enhancements in the elongated cases are weaker globally compared to the compressed cases.

\begin{figure}[htbp]
\suppressfloats
\begin{center}
\includegraphics[width=0.6\textwidth]{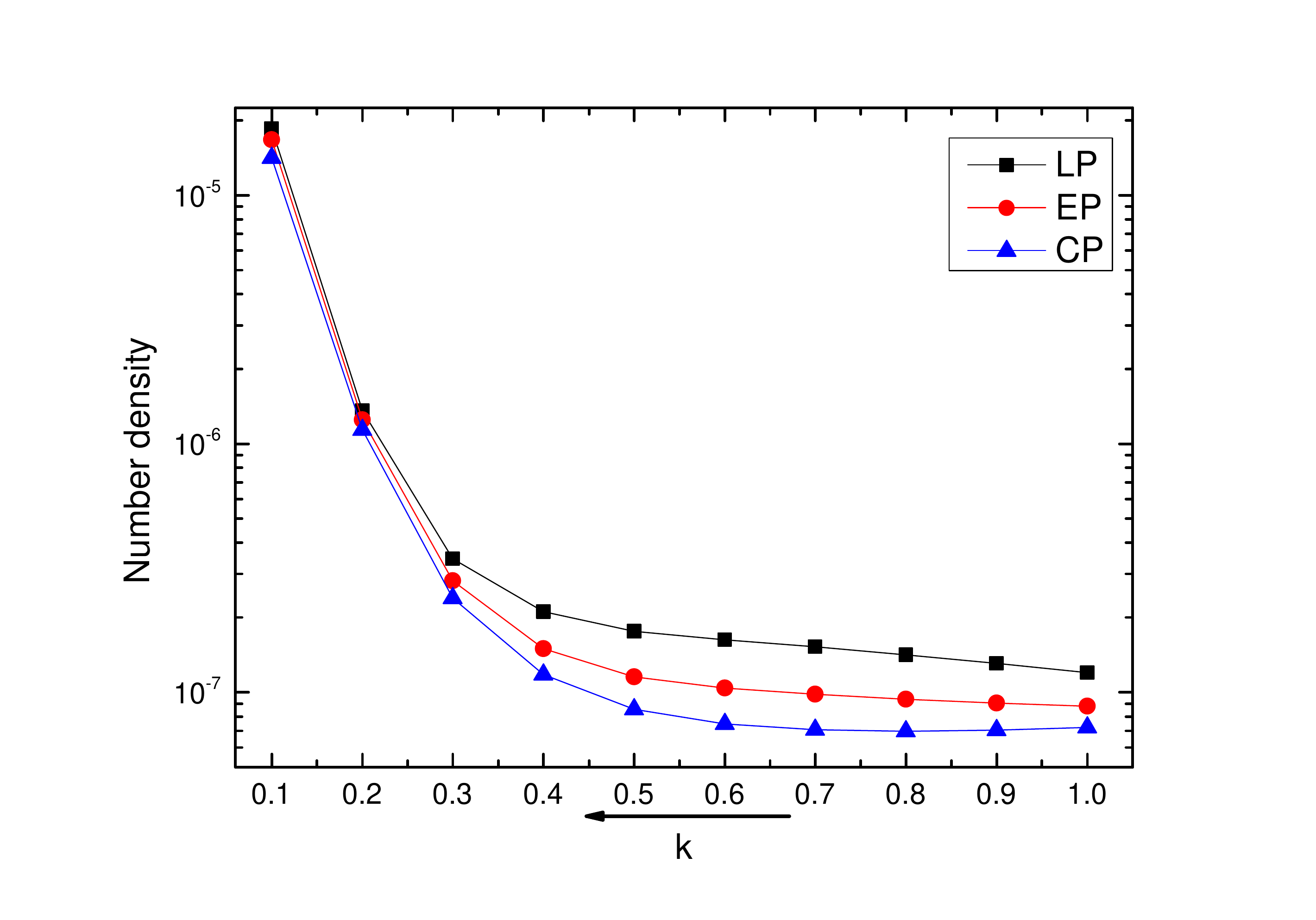}
\end{center}
\caption{The number density (in unit of $\lambda_{c}^{- 3} = m ^ 3$) of pairs produced in different polarized electric fields for the shorten falling length of asymmetric pulse shape with $0 < k \leq 1$. The field parameters are the same as in Fig.\ref{fig:1}. LP, EP and CP with squares, circles and triangles stands for linear $\delta = 0$, elliptical $\delta = 0.5$, and circular $\delta = 1$ case, respectively.}
\label{fig:7}
\end{figure}

\begin{figure}[htbp]
\suppressfloats
\begin{center}
\includegraphics[width=0.6\textwidth]{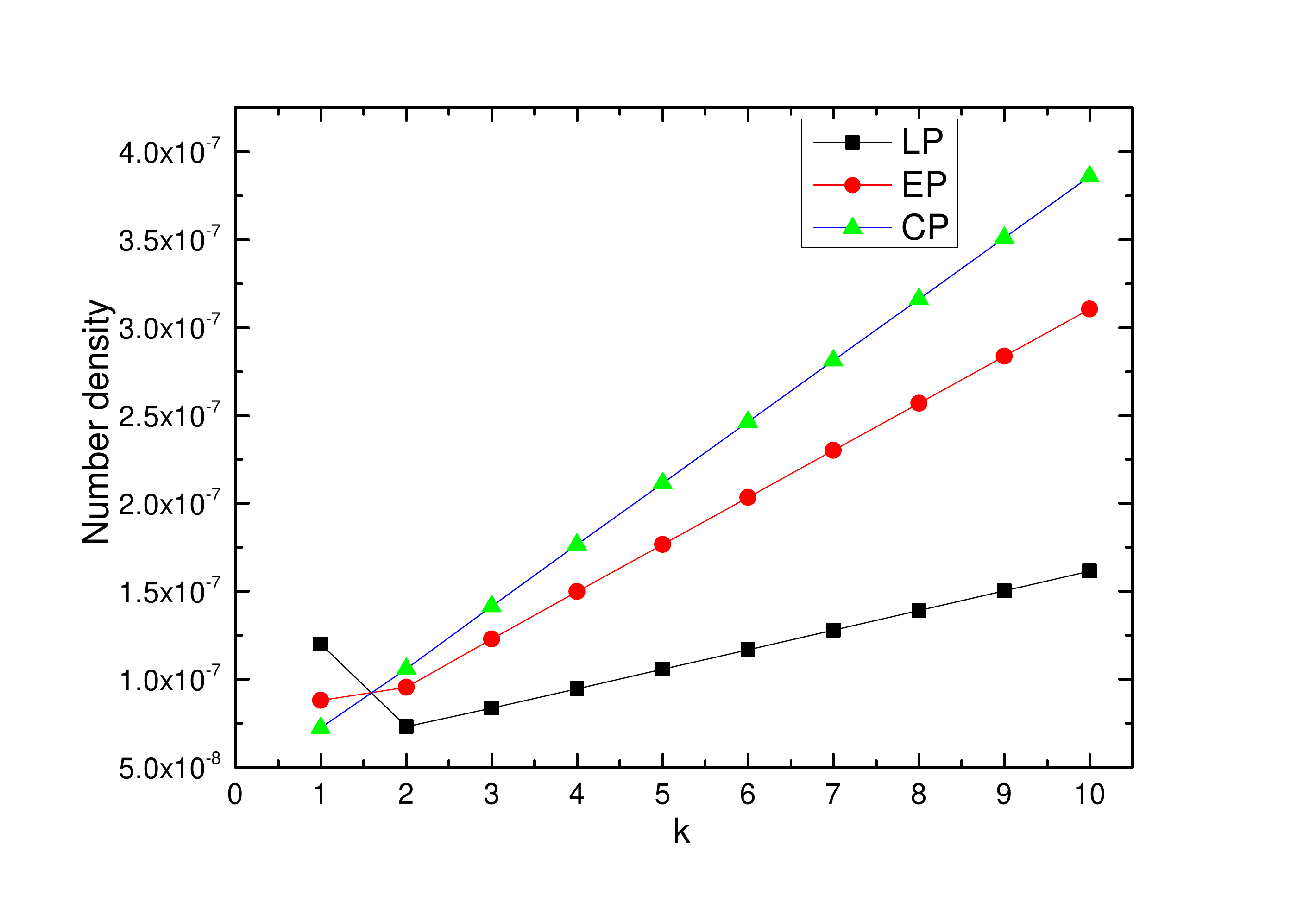}
\end{center}
\caption{Same as in Fig.\ref{fig:7} except for the elongated falling case with $k \geq 1$.}
\label{fig:8}
\end{figure}

\section{Number density of pair production}\label{number}

In this section, we calculate the change of the pair number density generated in different polarization electric fields with asymmetric shape and different pulse length ratio $k$. The results are shown in Figs. \ref{fig:7} and \ref{fig:8} for compressed and elongated falling pulse, respectively.

It is found that when the falling pulse width is compressed, i.e., $0< k\leq1$, the number density of created pairs decreases with the increase of electric field polarization. We find that the number density of electron-positron pairs in different polarizations increases with the decrease of pulse length ratio value $k $. For the larger compression it is more obvious especially for those of $k=0.4 $ to $k=0.1 $. When the pulse length is compressed, the number density increases by more than two orders of magnitude for each polarization.

Concretely, for the linear polarization the number density increases from $1.20 \times 10^{- 7}$ when $k = 1$ to $1.853 \times 10 ^{- 5}$ when $k = 0.1 $. For the elliptical polarization it increases from $8.799 \times 10^{- 8}$ when $k = 1 $ to $1.673 \times 10^{- 5}$ when $k = 0.1 $. For the circle polarization it increases from $7.237 \times 10^{- 8}$ when $k = 1 $ to $1.414 \times 10^{- 5}$ when $k = 0.1 $.

On the other hand, it can be also found that in the situation of the falling pulse elongated, the number density of created pairs is almost increasing with the field polarization parameter $\delta$ as well as the pulse elongation parameter $k$ except that for the linear polarization it has a little decrease when the falling pulse elongation is not large but then it still increases with the $k$ becoming larger and larger. This is mainly attributed effects of pulse length on the pair production processes. For the linear polarized electric field this pattern is also found in Ref \cite {Kohlfurst:2013}, where the authors have considered the single Sauter pulse. It is found that the particle number increases first with the increasing of pulse length until it reaches $\tau=0.5 m^{-1}$, then it decreases and reaches its minimum at $\tau=30m^{-1}$ and finally it increases again slowly, refer to Fig.4 of Ref \cite {Kohlfurst:2013}).

From Figs. \ref{fig:7} and \ref{fig:8}, one can infer that the number density exhibits polarization dependence for compressed pulse asymmetry and elongated pulse asymmetry of the field. For compressed pulse asymmetry cases, the number density decreases with the increase of the field polarization, while for the elongated pulse cases, the number density increases with the increase of the field polarization except the case of $k=1$. The reason is when $k=1$, the number of oscillation cycles of within the envelope pulse is $6$, and clean multiphoton pair production signal is not obvious. For compressed pulse cases $k<1$,  the standard multiphoton pair production is even less significant (for the small $\tau$, although the Keldysh parameter $\gamma= m/eE_{0}\tau>1$, but strictly it is not complete multiphoton process). For the elongation pulse $k>1$, as $k$ increases, pair creation is dominated by multiphoton mechanism, at this time for $\omega=0.6m$, the corresponding number density for the circular polarization is greater than that for the middle elliptical polarization, and the latter is greater than that for linear polarization cases (see also Fig.4 in Ref \cite {Li:2015cea} for a reference).

In a word, when the falling pulse length is compressed, the number density can be increased by two orders of magnitude, however, for the opposite case, i.e., when the falling pulse length is extended, the number density is enhanced only within the half orders of magnitude. Therefore, for asymmetric electric fields with different polarizations, in order to increase effectively the number density of the produced electron-positron pairs it is better to shorten the falling pulse. Note that in our previous work for linear polarized cases \cite {oluk}, where it is studied by solving the quantum Vlasov equation approach, the similar  finding has been presented qualitatively.

\section{Summery and discussion}

In this study, we investigated the effects of asymmetric pulse shape on the momentum
spectrum of created electron-positron pairs in strong electric fields for different polarization scenarios, in three different situations of linear, middle elliptical and circular polarized fields on the momentum spectrum of created particles by applying the DHW formalism. The main results for the spectra of produced pairs can be summarized as follows.

When the falling pulse length is compressed, for linear polarization the produced pairs spectra exhibit a shift and split of peaks. For middle elliptic polarization as well as circular polarization the momentum spectrum gets distorted and exhibits shift of peaks. Finally for each different polarization the peaks shifted to the central region at the momentum plane, therefor peak values enhanced two orders magnitude compared to the symmetric situation.When the falling pulse length is elongated, the ring structures appear for different polarization. It is also noted that for this asymmetric situation peak values increase with the field polarizations compared to the symmetric case while it is smaller  than that in compressed cases. Some phenomena of the momentum spectra are consistent with the effect of frequency chirp of our previous study \cite{olugh1}.

We also study the effect of asymmetric falling pulse on the obtained number density. It is found that the number density decreases or/and increases with the polarization for compressed or/and elongated falling pulse. It is important that, when the falling pulse length is compressed, the number density of the produced pairs can be enhanced significantly more than $2 $ orders of magnitude.

The results are helpful to understand the influence of pulse length, which is an important parameter of the external field, and to deepen the understanding of the external pulse structure. Although these results reveal some useful information about the production of $e^{+}e^{-}$ pairs in different elliptical polarization cases, in this study we restricted ourselves to the multiphoton pair creation, so the asymmetric pulse shape effects for pair creation under the Schwinger mechanism needs to be studied further for different polarized field.

The other important phenomena observed in our numerical results are the spiral structure in momentum spectra which has an intrinsic connection with the spin or/and orbital angular momentum of field photons as well as the produced electron-positron particles. The theoretical analysis for this characteristic is not easy and almost ignored completely in present study. However its abundant information about the rotation degree is very important and helpful to understand the involved strong external field interaction with vacuum and the possible application to the future real experiment.

\begin{acknowledgments}


\noindent
The work of O Olugh and BS Xie is supported by the National Natural Science Foundation of China (NSFC) under Grants No. 11875007 and No. 11935008. The work of ZL Li is supported by NSFC under Grant No. 11705278. The computation was carried out at the HSCC of the Beijing Normal University.

\end{acknowledgments}

\end{document}